\documentclass[twocolumn,secnumarabic, amssymb, superscriptaddress, nobibnotes, aps, pra]{revtex4-2}

\usepackage{amsmath,amssymb,mathtools}
\usepackage{stmaryrd}
\usepackage{times}
\usepackage[euler]{textgreek}
\usepackage[bbgreekl]{mathbbol}
\usepackage{graphicx}
\usepackage{bm,color}
\usepackage{natbib,hyperref}
\usepackage[capitalize]{cleveref}
\usepackage{multirow}

\hypersetup{colorlinks=true,linkcolor=blue,citecolor=blue,urlcolor=blue}

\begin{document}
	
	\title{Optical frequency combs significantly spanned to broad bandwidths by an optomechanical resonance}
	\author{Xin Gu}
	\affiliation{Fujian Key Laboratory of Light Propagation and Transformation \& Institute of Systems Science, 
		College of Information Science and Engineering, Huaqiao University, Xiamen 361021, China}
	\author{Jinlian Zhang}
	\affiliation{Fujian Key Laboratory of Light Propagation and Transformation \& Institute of Systems Science, 
		College of Information Science and Engineering, Huaqiao University, Xiamen 361021, China}
	\author{Shulin Ding}
	\affiliation{National Laboratory of Solid State Microstructures and College of Engineering and Applied Sciences, Nanjing University, Nanjing 210093, China}
		\author{Xiaoshun Jiang}
	\affiliation{National Laboratory of Solid State Microstructures and College of Engineering and Applied Sciences, Nanjing University, Nanjing 210093, China}
		\author{Bing He}
	\email{bing.he@umayor.cl}
	\affiliation{3Multidisciplinary Center for Physics, Universidad Mayor, Camino La Pir\'{a}mide 5750, Huechuraba, Chile}
	\author{Qing Lin}
	\email{qlin@hqu.edu.cn}
	\affiliation{Fujian Key Laboratory of Light Propagation and Transformation \& Institute of Systems Science, 
		College of Information Science and Engineering, Huaqiao University, Xiamen 361021, China}
	
	\begin{abstract}
	Optical frequency comb, as a spectrum made of discrete and equally
	spaced spectral lines, is a light source with essential applications in modern
	technology. Cavity optomechanical systems were found to be a feasible candidate for
	realizing on-chip frequency comb with low repetition rate. However, it was difficult
	to increase the comb line numbers of this type of frequency combs because the
	mechanical oscillation amplitude of such system, which determines the frequency
	comb bandwidth, cannot quickly increase with pump laser power. Here, we
	develop a new approach to generate broadband optomechanical frequency comb by
	employing a different mechanism to enhance the mechanical oscillation. Two
	pump tones with their frequency difference matching the mechanical frequency will
	drive the system into a self-organized nonlinear resonance and thus tremendously
	transfer the energy to the mechanical resonator. As a result, more than $10000$ or
	even more comb lines become available under the pump laser power in the order of
	milliwatt. A unique feature of the self-organized resonance is the mechanical
	frequency locking so that, within a certain range of the frequency difference between two drive tones, 
	the distance between comb teeth can be locked by the two drive tones and becomes independent of any 
	change of pump power. This property guarantees a stable repetition rate of the generated frequency comb.
	\end{abstract}
	
	\maketitle

\section{Introduction}
Optical frequency combs, a light source of a series of discrete and equally spaced frequencies, have found wide applications in modern technology, such as optical clock \cite{clock}, astronomical spectrograph calibration \cite{asc}, precision time/frequency transfer \cite{ptt}, ultraviolet and infrared (IR) spectroscopy \cite{uis1, uis2, uis3}, precision distance measurement \cite{pdm}, optical communication  \cite{oc}, and others. In the past, the main technologies developed to generate optical frequency combs are confined to mode-locked femtosecond lasers \cite{fs1, fs2}, electro-optic modulation of a continuous-wave laser \cite{eo}, and parametric frequency conversion through cascaded four-wave mixing \cite{OFCm1, OFCm2, pfc1, pfc2, pfc3}. Optomechanical nonlinearity in a micro-cavity \cite{ofc1,jiang,ofc4}, as well as the related magnotic realization \cite{ofc3}, is a recently interested approach to realize such light source known as optomechanic-optical frequency comb (OMOFC). OMOFCs are repeated pulses at the corresponding mechanical frequencies, which are relatively low, e.g, less than 1 GHz, and such low-repeating OMOFCs are highly relevant to some states-of-art technologies like high-resolution spectroscopy \cite{hr1, hr2}, spectral measurements \cite{sen}, multiphoton entangled states \cite{elr}, and high-power continua in holey fibers \cite{bc}.

A unique character of optomechanical nonlinearity is the generation of high-order sidebands, the field frequency components distanced exactly by the mechanical oscillation frequency, due to the energy transfer between the cavity field modes and the mechanical mode of a cavity optomechanical system (OMS) \cite{ofc1, jiang, step}. For practical applications OMOFCs must have more sidebands.  
However, the comb lines of the previously achieved OMOFCs are rather limited, only around dozens or hundreds as reported by Refs. \cite{ofc1, ofc4, ofc3}. A much more improved setup combining thermal-optic effect can make $938$ comb lines at a laser power of $448$ mW \cite{jiang}. For many applications such as precision optical metrology, such limited numbers of comb lines are not enough. A straightforward method for broadening comb span is to use higher pump power. However, the OMS can enter the regimes of chaos under very strong drive intensity, and the tremendously increased intracavity photons would damage the micro-cavity after turning on the strong pump laser for a period of time. An effective way to beat the bottleneck is crucial to the next stage of developing practical OMOFCs.  

As we will see later, the frequency span of an OMOFC is determined by how large the associated mechanical oscillation amplitude can be realized. Therefore, the more energy is transferred to the mechanical resonator, the wider the output cavity field spectrum will be. Here, we propose an approach of efficiently adding up the mechanical energy based on a newly discovered resonance mechanism \cite{step}: the mechanical oscillation of an unresolved-sideband OMS will be significantly enhanced under a two-tone driving field, with the difference of its tone frequencies being matched to the intrinsic mechanical frequency of the OMS. In this scenario the pump power 
for generating OMOFCs can be reduced to the level of only $\mu$W for obtaining dozens of comb lines. On the level of milliwatt (mW), more than $10000$ comb lines can be generated with the currently available technologies. A system optimized further can realize the frequency combs with even broader spans. 

\begin{figure}[ht]
	\centering
	\includegraphics[width=\linewidth]{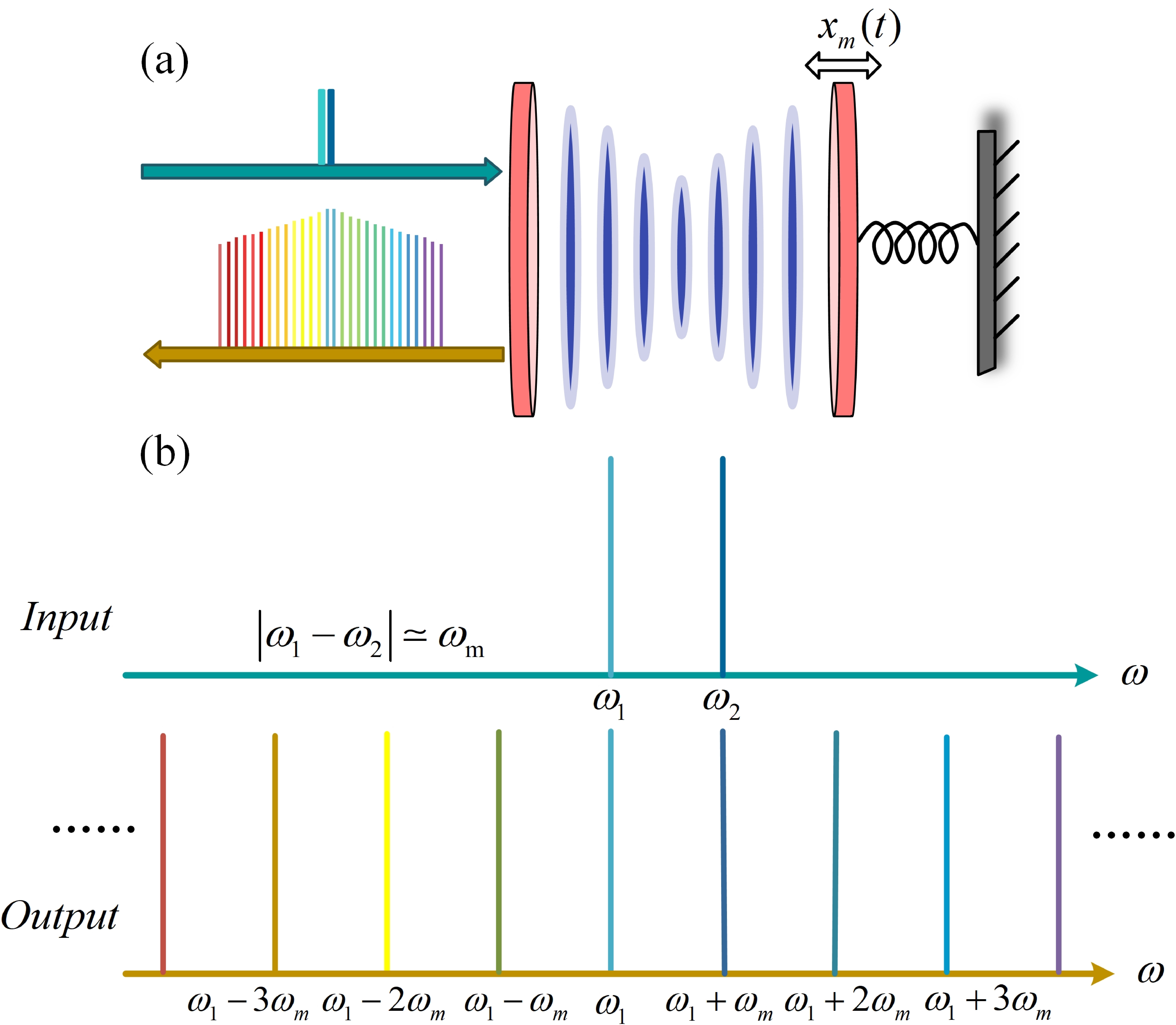}
	\caption{Generic setup for generating optomechanical frequency combs. (a) A two-tone pump laser field drives the optical cavity with a movable mirror as the mechanical resonator. If the frequency difference $|\omega_1-\omega_2|$ of the two tones gets close to the mechanical frequency $\omega_m$, a special nonlinear resonance of optomechanics can be induced, leading to a significantly enhanced mechanical oscillation and the corresponding broadband cavity field as the frequency comb. (b) Typical spectrum of the driving field and the intracavity field with high-order sidebands generated by the consecutive interaction with the mechanical mode, which is quantitatively described by Eq. (\ref{potential}).}
	\label{fig1}
\end{figure}

Another important issue for frequency combs is their stability, i.e., to preserve the comb lines perfectly equidistant for a period as long as possible, for hours and even for days. In the past experiments optical spring effect is unavoidable, depending on the drive power and resulting in unpredictable mechanical frequency shift that changes the comb line distance. Our adopted nonlinear resonance also has a property of mechanical frequency locking, rendering the comb line distance be locked to be close to the intrinsic mechanical frequency of the system, totally independent of the applied drive power. It can thus practically guarantee the stability of the generated OMOFCs. 

The rest of the paper is organized as follows. In Sec. \ref{mechanism}, the mechanism of realizing large mechanical oscillation by a special two-tone driving field is detailed in comparison with the corresponding mechanism due to a single-tone laser field, to illustrate the important features of broadband OMOFC and the associated oscillation frequency locking. The optimization of the concerned OMOFC by a parallel displacement of the drive tones is discussed in Sec. \ref{displace}. All factors affecting the performance of the system, including the drive intensity and intrinsic mechanical frequency are illustrated in Sec. \ref{performance}. Then the technical requirements for the used drive tones is sketched in Sec. \ref{requirement}, before the final part of conclusive remarks.

\section{Mechanism of generating broadband OMOFC}
\label{mechanism}

The system we consider is a generic OMS, which is represented by the Fabry-Perot cavity in Fig. \ref{fig1}: the movable mirror acts as a mechanical resonator with frequency $\omega_m$ and the cavity is pumped by a two-tone laser field with their frequencies $\omega_1$ and $\omega_2$, respectively. The radiation pressure of the cavity field, which is proportional to the photon number $|a|^2=(X^2_c+P^2_c)/2$ ($X_c$ and $P_c$ are two quadratures of the field), drives the mechanical resonator into oscillation once the pump field satisfies a certain condition. As the first-order correction to the cavity field energy by the mechanical displacement under radiation pressure, there is the interaction potential
\begin{align}
	V_{int}=-g_mX_m|a|^2
	\label{potential}
\end{align}
for the cavity field and mechanical resonator, where $g_m$ is the optomechanical coupling strength at the single photon level. Here we have introduced the dimensionless mechanical displacement $X_m$ by the scaling $x_m=\sqrt{\frac{\hbar}{m\omega_m}}X_m$ ($m$ is the effective mass of the mechanical resonator), and the conjugate dimensionless momentum $P_m$ is defined by another scaling $p_m=\sqrt{m\hbar\omega_m}P_m$. 
Then the dynamical equations of the system are given in the rotation frame with respected to the cavity frequency $\omega_c$ as follows:
\begin{align}
	\dot{a}&=-\kappa a+ig_mX_ma+\sum_{n=1,2}E_ne^{i\Delta_n t},\nonumber\\
	\dot{X}_m&=\omega_mP_m,\nonumber\\
	\dot{P}_m&=-\omega_mX_m-\gamma_mP_m+g_m|a|^2+\sqrt{2\gamma_m}\xi_m(t),
	\label{dy}
\end{align}
where the detuning is defined as $\Delta_{1(2)}=\omega_c-\omega_{1(2)}$. The cavity damping rate $\kappa$ includes the coupling rate to the external field $\kappa_e$ and the intrinsic loss $\kappa_i$, so that $\kappa=\kappa_e+\kappa_i$. The drive amplitudes 
are determined by the pump power $P$ as $E_{1(2)}=\sqrt{2\kappa_e P/\hbar \omega_{1(2)}}$. Moreover, at the room temperature, the thermal noise can be treated as a white noise satisfying the correlation $\langle \xi_m(t)\xi_m(t')\rangle=(2n_{th}+1)\delta(t-t')$ \cite{T-entangle}, where $n_{th}=(e^{\hbar\omega_m/k_BT}-1)^{-1}$ is the thermal phonon number at the temperature $T$. This thermal noise can be simulated with a stochastic function \cite{noise}, but it does not have a significant effect on the concerned dynamical 
processes \cite{step}, so we will ignore it and adopt a mean-field approximation in the following discussions.

\begin{figure*}[tb]
	\centering
	\includegraphics[width=0.9\linewidth]{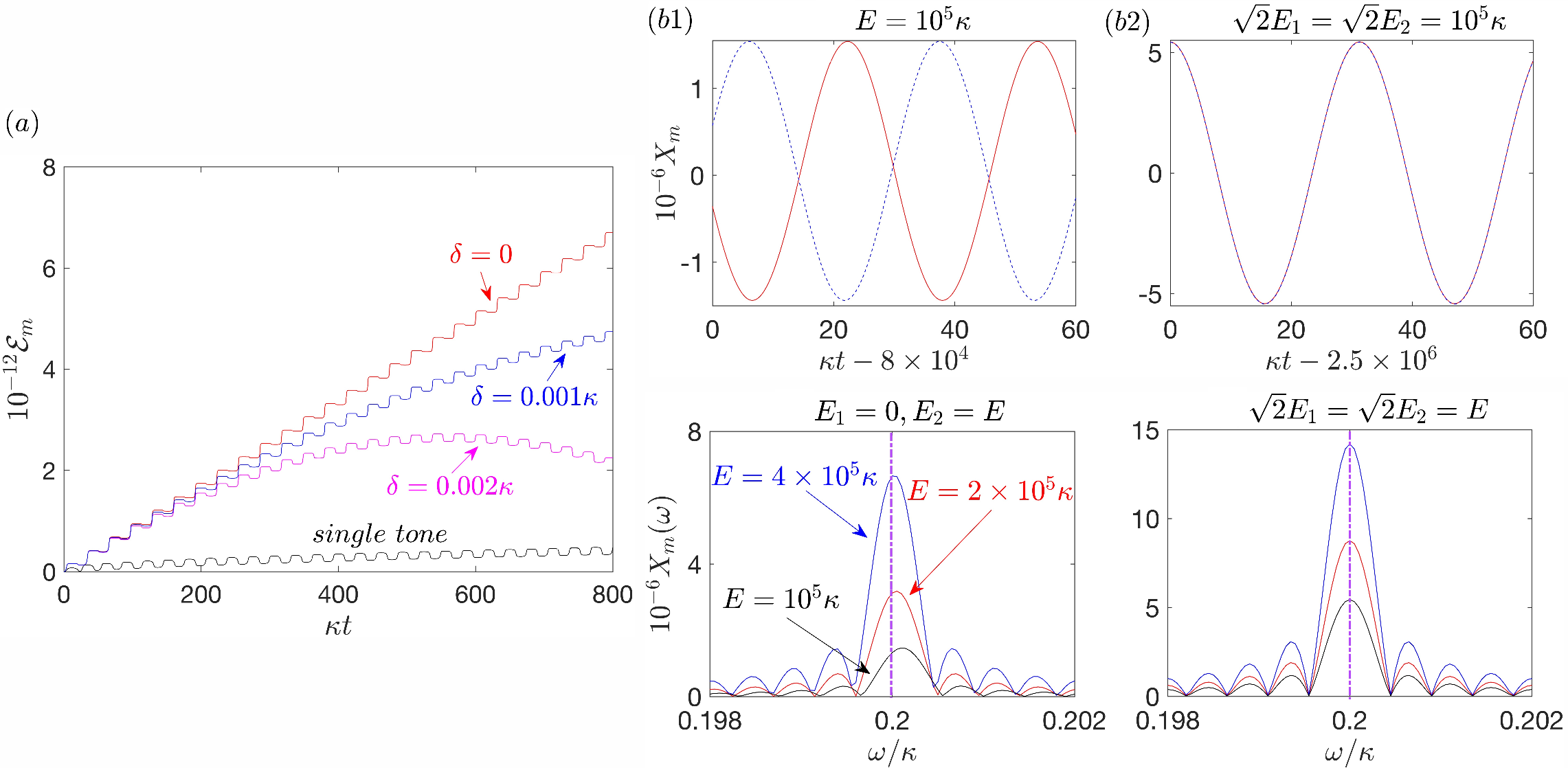}
	\caption{A self-organized optomechanical resonance and its associated mechanical frequency locking. (a) The emergence of the resonance under the condition $\delta=0$ in Eq. (\ref{fm}). The increased mechanical energy is at a rate much higher than the corresponding one under a single-tone pump of the same power. (b1)-(b2) The stabilized mechanical oscillation (the red solid curve) is compared with a reference oscillation at the frequency $\omega_m$ (the blue dashed curve)---the single-tone scenario shows their phase difference accumulated from a slight difference in their frequencies, but the two-tone scenario shows their exact overlap due to a frequency locking (the system has completely stabilized during the illustrated period of time, as the evolution time has been much larger than $\gamma_m^{-1}$). Moreover, the mechanical frequency shifts due to optical spring effect are obvious in the scenario of single-tone pump, in contrast to the completely locked mechanical frequency at the intrinsic one $\omega_m$ by the two-tone pump (the peaks of mechanical spectrum for all different drive amplitudes $E$ are exactly at the point $\omega=\omega_m$). In the unresolved sideband regime where the stabilized cavity field is pulsed, a higher pump power gives rise to less spring effect for the system driven by a single-tone pump, because the cavity field pulse becomes narrower. The used parameters are $g_m=10^{-5}\kappa, \omega_m=0.2\kappa, \gamma_m=10^{-5}\kappa$. In (a) the drive amplitude is $\sqrt{2}E=10^5\kappa$.}
	\label{fig2}
\end{figure*}

\subsection{Field sidebands due to a single-tone drive}

First, we take a look at the scenario when an OMS is driven by a single-tone pump field. When the power of the pump field is higher than a certain threshold of Hopf bifurcation, the system will enter the coupled oscillations of cavity field and mechanical resonator.
The mechanical oscillation generally stabilizes in the form 
\begin{align}
	X_m(t)=A_m\cos(\Omega_mt)+d_m,
	\label{m}
\end{align}
where $A_m$ denotes the oscillation amplitude and $d_m$ a pure displacement with $d_m\ll A_m$. It should be noted that, due to the existence of optical spring effect, the frequency $\Omega_m$ can considerably differ from the intrinsic mechanical frequency $\omega_m$, which is determined by the fabrication of the system. If one plugs this $X_m(t)$ into the first equation of Eq. (\ref{dy}), which is reduced to the situation of a single-tone drive, there will be 
the evolved cavity field in the form $a(t)=\sum_n a_ne^{in\Omega_mt}$ (up to a global phase), where
\begin{align}
	a_n=E\sum_{p}\frac{J_{n-p}(g_mA_m/\Omega_m)J_{p}(-g_mA_m/\Omega_m)}{ip\Omega_m+\kappa+i(g_md_m-\Delta)},
	\label{term}
\end{align}
with $J_p(x)$ being the Bessel function of the first kind (see the derivation in Appendix). All these components $a_n=|a_n|e^{i\varphi_n}$ are called sidebands of the cavity field, from which it can be seen that their phases $\varphi_n$ 
are fixed under a set of given system parameters, but the different sidebands have their different oscillation phases. The variable $g_mA_m/\omega_m$ of the Bessel functions is highly relevant to the sideband structure. There is one property of the Bessel function: given a fixed $x$, the magnitude of the $p$-th order Bessel function $J_p(x)$ will decay quickly if $|p|>|x|$; this property also guarantees the convergence of the summation in Eq. (\ref{term}). Then, the value of $g_mA_m/\Omega_m$ determines to which sideband $p$ the cavity field can have. In other words, the condition $p<g_mA_m/\Omega_m$ or $-p>-g_mA_m/\Omega_m$ specifies a range
\begin{align}
	-g_mA_m<\Delta \omega <g_mA_m
	\label{range}
\end{align}
covering all possible sidebands with their frequency $p\Omega_m$ ($p\in Z$) in a reference frame rotating at the pump laser frequency. The argument $g_mA_m/\Omega_m$ of the Bessel functions in Eq. (\ref{term}) happens to be the half of the comb line number of the generated cavity field as frequency comb (the extension to the negative integers $p$ gives the other half). 
Broad comb span should be certainly obtained by having a large mechanical amplitude $A_m$. The straightforward way for having a large $A_m$ is to enhance the drive amplitude $E$, but the drive power cannot be increased without a limit because the accompanying heat would damage the systems. As reported by a recent experiment \cite{jiang}, a large pump detuning to the blue-detuned side can also increase the mechanical amplitude due to a peculiar system dynamics of OMS (see the discussions in Sec. 3). Another approach for having more sidebands is to make of the unresolved sideband regime ($\omega_m<\kappa$), where pulsed cavity field can be generated \cite{s1,s2,s3}. It is easy to reach the unresolved sideband regime by adjusting the pump-field coupling rate $\kappa_e$ for a system with fixed intrinsic mechanical frequency $\omega_m$. In what follows, we will consider this unresolved sideband regime for the designs of OMOFCs.

\begin{figure*}[ht]
	\centering
	\includegraphics[width=0.8\linewidth]{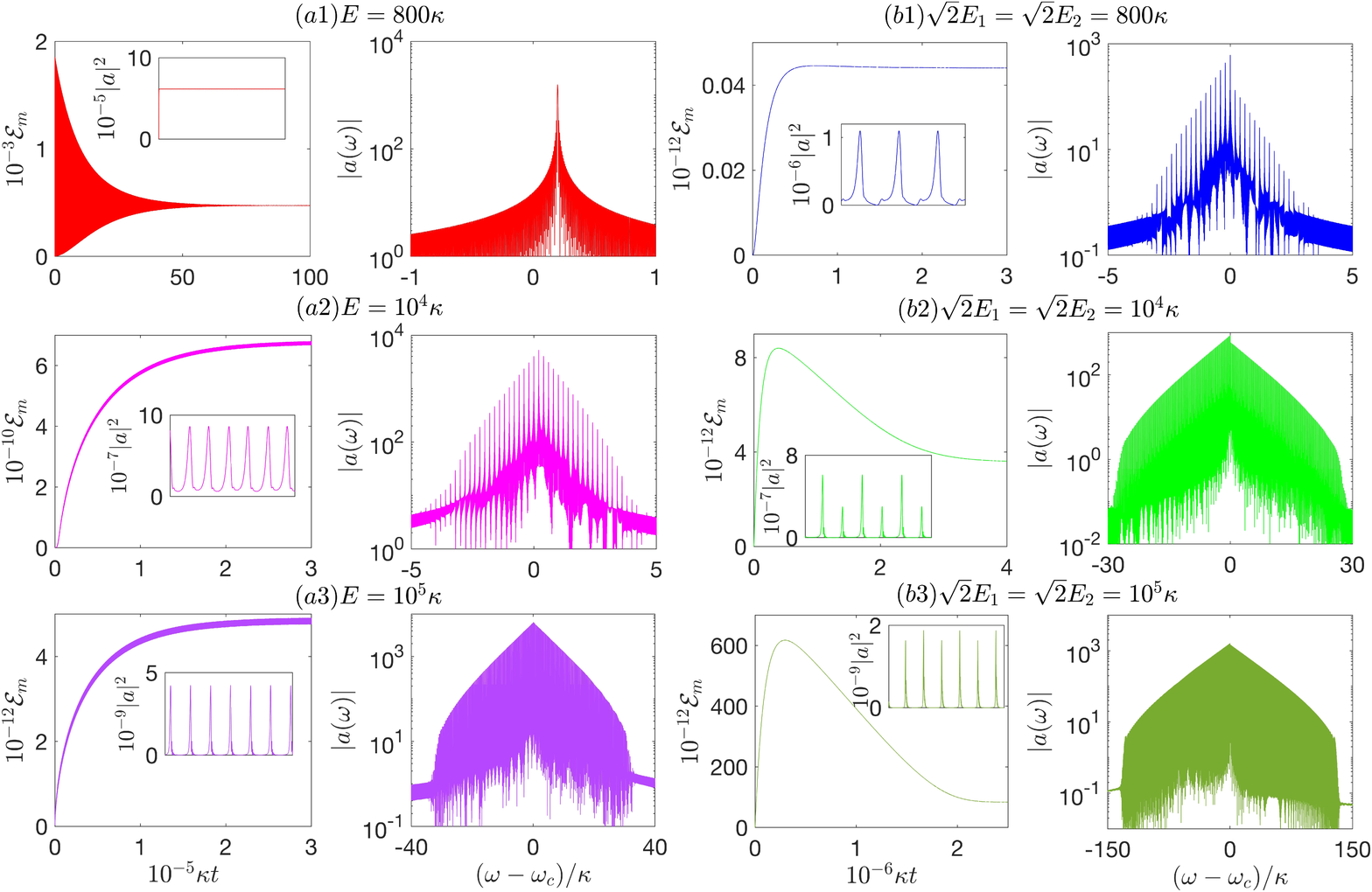}
	\caption{Comparisons of the stabilized OMOFCs generated by a single-tone field at $\Delta=-\omega_m$ and a two-tone field with $\Delta_1=-\omega_m$ and $\Delta_2=0$. Given the same pump power, the comb span of the latter is obviously much wider. The inset 
		in each panel shows the stabilized cavity field intensity corresponding to the stabilized mechanical energy $\mathcal{E}_m$. As the pump amplitude $E$ is increased, the illustrated field pulses become narrower, having a broader span of the generated OMOFC. The system parameters are chosen as $g_m=10^{-5}\kappa, \omega_m=0.2\kappa$, and $\gamma_m=10^{-5}\kappa$.}
	\label{fig3}
\end{figure*}

\subsection{Enhanced optomechanical oscillation by a special two-tone pump}
The sideband amplitude in Eq. (\ref{term}) is based on an already realized mechanical amplitude $A_m$, which is through a complicated dynamical process that must consider the coupled nonlinear equations Eq. (\ref{dy}). All similar processes vary with the applied pump field. If the single-tone pump is replaced by a two-tone driving field, the stabilized mechanical amplitude will have totally different patterns. The two tones with their corresponding detunings $\Delta_1$ and $\Delta_2$ satisfy the relation:
\begin{align}
	|\Delta_1-\Delta_2|=\omega_m+\delta,
	\label{fm}
\end{align}
where the deviation $\delta$ can be any real number. Once the above difference between the two tones gets close to the intrinsic mechanical frequency $\omega_m$, i.e. $\delta\rightarrow 0$, the system being operated in the unresolved sideband regime ($\omega_m<\kappa$) will exhibit a resonance phenomenon shown in Fig. \ref{fig2}(a). Generally, various nonlinear resonances can 
occur when a particular nonlinear oscillator has the maximum response to an external drive \cite{resonance}. Our concerned resonance is out of a self-organization of the system when it is operated under the two conditions, $|\Delta_1-\Delta_2|=\omega_m$ and $\omega_m/\kappa<1$ \cite{step}. In the unresolved sideband regime satisfying $\omega_m/\kappa<1$, 
two cavity field pulses are generated in each mechanical oscillation period; one of them emerges when the mechanical 
resonator approaches its positive top speed in the oscillation period, but the other manifests when it is close 
to its negative top speed (the radiation force of the pulse is opposite to the moving direction of the mechanical oscillator). Under a further condition $\delta=0$, the field pulse acting in the opposite direction of the mechanical resonator's movement will be significantly suppressed, so the other pulse acting along with the movement of the mechanical resonator will successively add more and more energy 
\begin{align}
	\mathcal{E}_m=\frac{1}{2}X_m^2+\frac{1}{2}P_m^2\approx \frac{1}{2}A_m^2	
	\label{em}
\end{align}
to it, in the form of a quasi-linear energy increase in Fig. \ref{fig2}(a). Such tremendously increased mechanical amplitude $A_m$ surely gives rise to a broadband $2g_mA_m$ of the cavity field, since, according to the reduced dynamics of the cavity field part [by plugging the stabilized mechanical oscillation $X_m(t)$ into the first equation of Eq. (\ref{dy})], the stabilized mechanical amplitude $A_m$ still determines the cavity field frequency span so that the range in Eq. (\ref{range}) is valid (such sideband ranges can be accurately verified by the numerical calculations with our concerned two-tone drives). More features of the used nonlinear resonance 
are detailed in Ref. \cite{step}. The main point of the current work is how to utilize this mechanism to realize good-quality OMOFCs.

In the resolved sideband regime $\omega_m\gg \kappa$, a two-tone field satisfying the condition $\delta=0$ realizes a complete locking of an OMS: the mechanical oscillation amplitude, the mechanical oscillation frequency, as well as the oscillation phase, can be fixed without varying even after a considerable change of the pump power and the associated noise effects, once the system has stabilized \cite{level1,level2}. These properties can be applied for various precise sensing \cite{mass,df1,df2}. The system operating in the unresolved sideband regime of $\omega_m/\kappa<1$ also inherits the property of frequency locking. Due to the existence of optical spring effect, there is an unavoidable and mostly obvious shift of the actual mechanical oscillation frequency $\Omega_m$ from the intrinsic mechanical frequency $\omega_m$ determined by system fabrication, when the system is driven by a single-tone field. An intuitive view of the effect is that the difference between the two frequencies, $\Omega_m$ and $\omega_m$, will accumulate a phase difference between the actual mechanical oscillation and a reference mechanical oscillation at the frequency $\omega_m$, after the oscillations proceed for a period of time; see Fig. \ref{fig2}(b1). This effect causes a problem of instability of frequency combs' repetition rates, as it was observed by the measurement of the Allan deviations with  changed pump powers \cite{jiang}. However, if the same single-tone pump field is split into two components with the difference of their frequencies matching the intrinsic mechanical frequency $\omega_m$, the stabilized mechanical oscillation will well keep its frequency to the intrinsic one $\omega_m$, as seen from the exactly coinciding actual and reference oscillation in Fig. \ref{fig2}(b2). It is even clearer to see the oscillation frequency locking by comparing the mechanical spectrum induced by the single-tone and double-tone pump, which are in the lower panel of Fig. \ref{fig2}(b1) and Fig. \ref{fig2}(b2), respectively. More generally, the mechanical oscillation frequency is locked to the frequency difference $|\Delta_1-\Delta_2|$ of two drive tones, as long as the error $\delta$ in Eq. (\ref{fm}) is within a certain range. Such a frequency locking is independent of the applied pump power and totally robust against noise perturbations.

\subsection{OMOFC broadened with an optomechanical resonance}

The advantages of the two-tone pump satisfying $|\Delta_1-\Delta_2|=\omega_m$ are clearer by a comparison with the corresponding single-tone pump; see Fig. \ref{fig3}. In the right part of Fig. \ref{fig3}, the corresponding pumps of a single tone (in the left part) are divided into two with one of them being shifted to a different frequency of $\Delta_2=0$. Dozens of comb lines can be obtained with a rather low pump power proportional to $E^2$, while the system driven by a single-tone drive below its Hopf bifurcation point stabilizes to a static equilibrium; compare Figs. \ref{fig3}(a1) and \ref{fig3}(b1). The comb span in Fig. \ref{fig2}(b1) is close to that in Fig. \ref{fig2}(a2), where the pump power is, however, increased by $156$ times. In Fig. \ref{fig3}(b3) the number of comb lines has been beyond $1000$ and, given the experimental setup in Ref. \cite{jiang}, this OMOFC can be achieved with a pump power below $100$ mW. Because we split an original pump into two tones, the corresponding comb teeth brightness will be lower due to the energy conservation. The brightness will be improved if both pump tones have the same power as in the corresponding single-tone scenario. For example, given the pump field of $E_1=E_2=10^4\kappa$ for the process in Fig. \ref{fig3}(b2), the amplitude of the highest comb tooth will be increased to $5.21\times 10^3$, well close to the corresponding one with $5.28\times 10^3$ in Fig. \ref{fig3}(a2).

\begin{figure}[h!]
	\centering
	\includegraphics[width=0.9\linewidth]{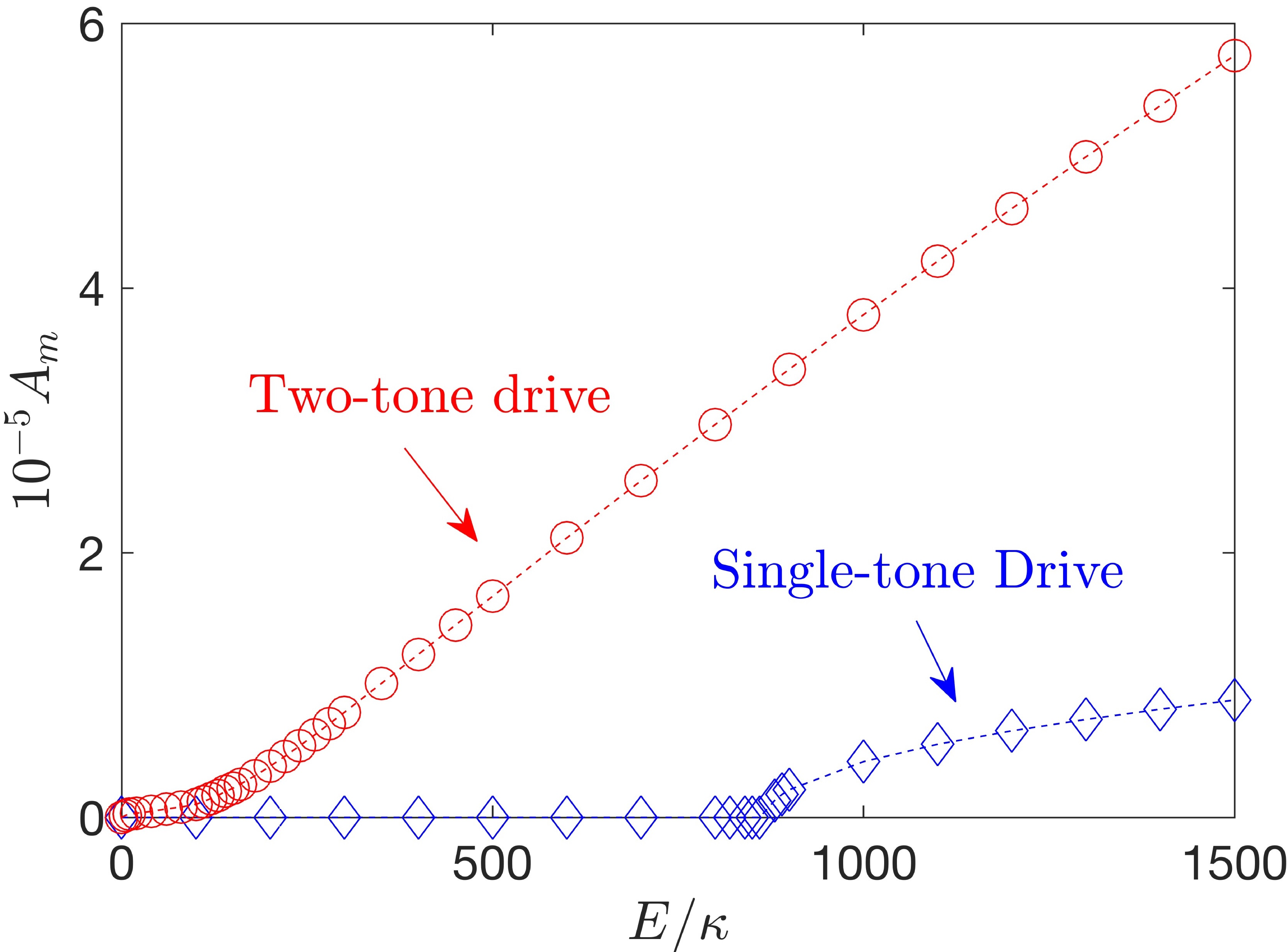}
	\caption{Comparison of the realized mechanical oscillation amplitudes by a single-tone pump and a corresponding two-tone pump. In the two-tone scenario the optomechanical oscillation exists with tiny amplitude near $E=0$. In the single-tone scenario, on the other hand, a Hopf bifurcation at $E\approx 880\kappa$ starts the oscillation from $A_m=0$, and it is evidenced by a critical slowing-down near the point (in Fig. \ref{fig3}(a1) the stabilization process close to the bifurcation has been rather slow). Here the system parameters are the same as those in Fig. \ref{fig3}.}
	\label{fig3b}
\end{figure}

Different patterns of adding up mechanical energy explain the difference in the generated OMOFCs. As seen from Figs. \ref{fig3}(a2) and \ref{fig3}(a3), the mechanical energy $\mathcal{E}_m$ under a single-tone pump increases in a manner of asymptotic stabilization with time. A stronger pump drives the same system to a higher mechanical energy, but the energy increase rate is not so significant. 
On the other hand, the system driven by a two-tone field satisfying $|\Delta_1-\Delta_2|=\omega_m$ exhibits a rapid growth of mechanical energy due to a self-organized resonance at the beginning stage of time evolution \cite{step}. Only after this stage, will the system gradually tend to a stabilization in a process dominated by its mechanical damping. The asymptotic behaviors of the systems operated in the unresolved sideband regime ($\omega_m<\kappa$) are very different from those in the resolved sideband regime ($\omega_m\gg\kappa$), where the optomechanical interaction enhances the mechanical damping so that the total effective damping rate is increased with an extra term $\Gamma_{opt}$ \cite{RMP}. Optomechanical cooling processes in the regime of resolved sideband are therefore much faster than the mechanical relaxation characterized by the time $\gamma_m^{-1}$ \cite{cooling,cool2,cool3}. In contrast, a cavity field realized in an unresolved sideband regime becomes pulsed, affecting the mechanical resonator only within a short period of time. Therefore, the intrinsic mechanical damping rate $\gamma_m$ determines the stabilization in the regime of $\omega_m<\kappa$. From Fig. \ref{fig3} one sees that the processes under a single-tone pump stabilize after a time in the order of $\gamma_m^{-1}$ but, due to the existence of a nonlinear resonance, the dynamical processes under the concerned two-tone drives are modified and take even longer time than $\gamma_m^{-1}$ when they become finally stabilized. An even clearer comparison between the single-tone scenario and our concerned two-tone scenario is displayed in Fig. \ref{fig3b}, where the achieved mechanical amplitudes $A_m$ under the lower drive amplitude $E$ have the totally different tendencies---there is a definite Hopf bifurcation point to start optomechanical oscillation for the single-tone scenario (a behavior called ``phonon laser" \cite{RMP}), but the optomechanical oscillation induced by the concerned two-tone pump has a continuous existence towards $E=0$ \cite{step}. The realized bandwidths $2g_mA_m$ of OMOFCs by the concerned two-tone pump fields, therefore, significantly surpass those by the corresponding single-tone pump fields.

The adopted exemplary system in Figs. \ref{fig3} and \ref{fig3b} has a mechanical quality factor $Q=2\times 10^4$. Due to the extremely slow stabilization processes caused by the optomechanical resonance, we have not simulated the dynamical evolution with a still smaller mechanical damping $\gamma_m$. OMOFCs covering much more sidebands are expected with the still improved mechanical quality factors. The continued discussions will be about how to achieve the better OMOFCs by the adjustment of other system parameters under the proper conditions.

\begin{figure}[h!]
	\centering
	\includegraphics[width=0.99\linewidth]{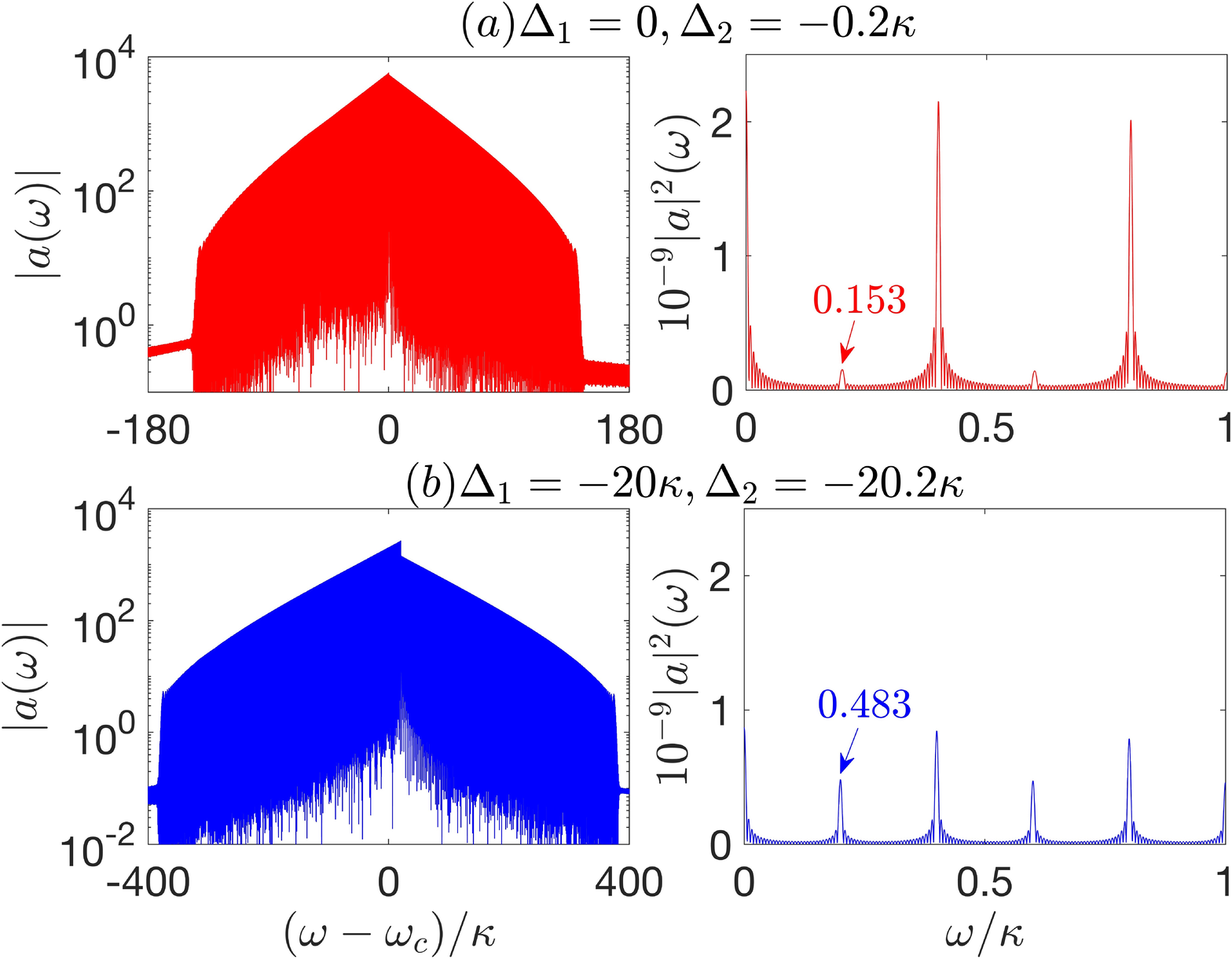}
	\caption{Illustration of the effect by a parallel displacement of two tones. From (a) to (b), both tones are displaced to more blue-detuned side by an amount of $20\kappa$, so that the comb span becomes broadened. The peak of the spectrum in (b) is shown to be shifted to $\omega=\omega_c+20\kappa$, and the amplitudes of all sidebands less than this peak frequency are enhanced. One result of the process is that the indicated first sideband magnitude $A_1$ on the right side of (a) is increased to the one on the right side of (b). Here the drive amplitude is $\sqrt{2}E=4\times10^5\kappa$, with all other parameters being the same as those in Fig. \ref{fig3}.}
	\label{fig4}
\end{figure}

\section{Optimization of OMOFC simply by detuning}
\label{displace}

Driven by a single-tone pump, there are two tendencies for a stabilized cavity field intensity, $|a(t)|^2=\sum_n A_n\cos(n\Omega_mt+\phi_n)$ ($\phi_n$ is the phase for the $n$-th sideband), with the increased blue detuning of the pump. One is certainly that the overall cavity field will dwindle as the pump is off the resonance. 
But the other is an energy redistribution among the field sidebands so that the first sideband of $|a(t)|^2$ with its magnitude, $A_1=2|\sum_p \alpha_{p+1}\alpha_p^{\ast}|$, will be enhanced. This first sideband with the magnitude $A_1$ predominantly determines the mechanical amplitude $A_m$ since it is the closest to the mechanical resonance frequency $\omega_m$ among all those with their magnitudes $A_n$, and an enhanced magnitude $A_1$ by blue detuning will thus increase the bandwidth $2g_mA_m$ of the generated OMOFC. These two tendencies compete with each other in the determination of the sideband magnitude $A_1$, and give rise to an optimum pump detuning for the bandwidth of OMOFC, beyond which the comb span will drop instead. An experimental demonstration of such enhancement by blue detuning is reported in Ref. \cite{jiang}, which shows the generation of the OMOFCs with about $1000$ comb lines by a large detuning further increased by a thermal-optic effect. 

When it comes to our concerned two-tone pump scenario, there is also a similar enhancement of comb span 
by varying drive tones. The previously mentioned optomechanical resonance still exists after the application the following displacements \cite{step}:
\begin{align}
	\Delta_1\rightarrow \Delta_1+\delta\omega, \ \  \Delta_2\rightarrow \Delta_2+\delta\omega,
\end{align}
for the two tones of the pump from their original frequencies of $\Delta_1=0,\ \Delta_2=-\omega_m$ we have adopted before. All features of the applied optomechanical resonance are well
preserved as long as the two tones satisfy the condition $|\Delta_1-\Delta_2|=\omega_m$. Now we demonstrate that an appropriate displacement $\delta \omega_m$ to the blue-detuned side can considerably broaden the comb span, as shown in Fig. \ref{fig4} where a displacement of $\delta\omega=-20\kappa$ extends the original comb span by about $2.64$ times.

\begin{figure}[h!]
	\centering
	\includegraphics[width=0.99\linewidth]{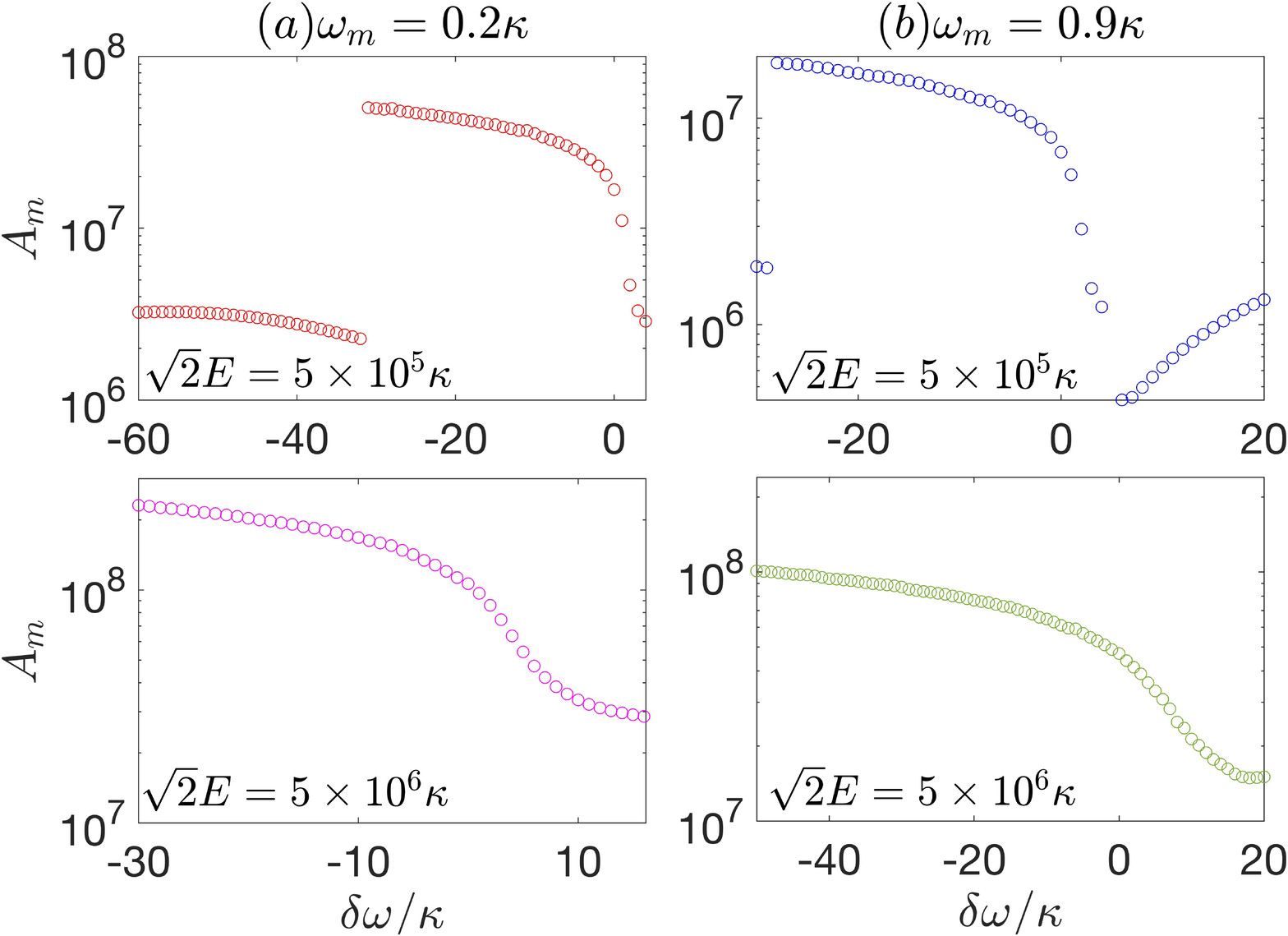}
	\caption{Induced mechanical amplitude $A_m$ under frequency displacement $\delta\omega$ of two drive tones for two different mechanical frequencies and two drive amplitudes. A suitable displacement can efficiently enhance the mechanical amplitude. In the two upper panels, the amplitude $A_m$ reaches the maximum at $\delta\omega=-32\kappa$ for $\omega_m=0.2\kappa$ and at $\delta\omega=-29\kappa$ for $\omega_m=0.9\kappa$. Given a higher pump power as in the lower panels, the possible displacement range can be extended further to reach an optimum point, for example, $\delta \omega=-137\kappa$ for $\omega_m=0.2 \kappa$ and $\delta\omega=-116\kappa$ for $\omega_m=0.9\kappa$ (the curves extending to the left sides will be broken and drop down at these points). The fixed parameters are the same as those in Fig. \ref{fig3}.}
	\label{fig5}
\end{figure}

Such spectrum broadening of OMOFC by a parallel displacement of the drive tones can be understood with a slight asymmetry of the comb spectrum shown in Fig. \ref{fig4}(b). The parallel displacement will move the peak of the optical spectrum from the original position at $\omega=\omega_c$ to a displaced position at $\omega=\omega_c+|\delta\omega|$, and the sidebands with their frequencies less than 
the original peak frequency will be enhanced after the displacement; see Fig. \ref{fig4}(b). This asymmetry amplifies the first sideband magnitude $A_1$ in the optical field intensity spectrum $|a|^2(\omega)$, as indicated by the values numerically obtained in Fig. \ref{fig4}. Like in a single-tone scenario \cite{jiang}, a larger mechanical amplitude $A_m$ due to the enhanced sideband magnitude $A_1$ will therefore lead to a further extended comb span $2g_mA_m$.

In Fig. \ref{fig5}, we show the relations between the dimensionless mechanical amplitude $A_m$ and 
the frequency displacement $\delta\omega$ for two different systems with their mechanical frequencies $\omega_m=0.2\kappa$ and $0.9\kappa$, which are also respectively driven by one two-tone pump of $\sqrt{2}E=5\times10^5\kappa$ and another two-tone pump of $\sqrt{2}E=5\times10^6\kappa$. The comb span can be broadened further by about three times for the lower drive power, while it can be broadened by about $2.2$ times for the higher drive amplitude. For the system described in Fig. \ref{fig3}, the comb line number will be more than $10000$ after the dimensionless mechanical amplitude is higher than $A_m=10^8$. In the upper panels of Fig. \ref{fig5}, such improvement will be stopped at where the curve of $A_m$ is broken at the blue detuning with $|\delta\omega|=32\kappa$ for $\omega_m=0.2\kappa$, as well as at $|\delta\omega|=29\kappa$ for $\omega_m=0.9\kappa$. These points indicate the optimum comb spans around there. The applicable displacement range can be increased further with a still larger drive amplitude $E$, as shown in the two lower panels of Fig. \ref{fig5}. Overall, the associated dynamical processes modified by drive-tone displacements are rather complicated due to the relevance of various factors.

\section{Performance with varied pump powers and due to different system fabrications}
\label{performance}

\begin{figure}[h]
	\centering
	\includegraphics[width=0.99\linewidth]{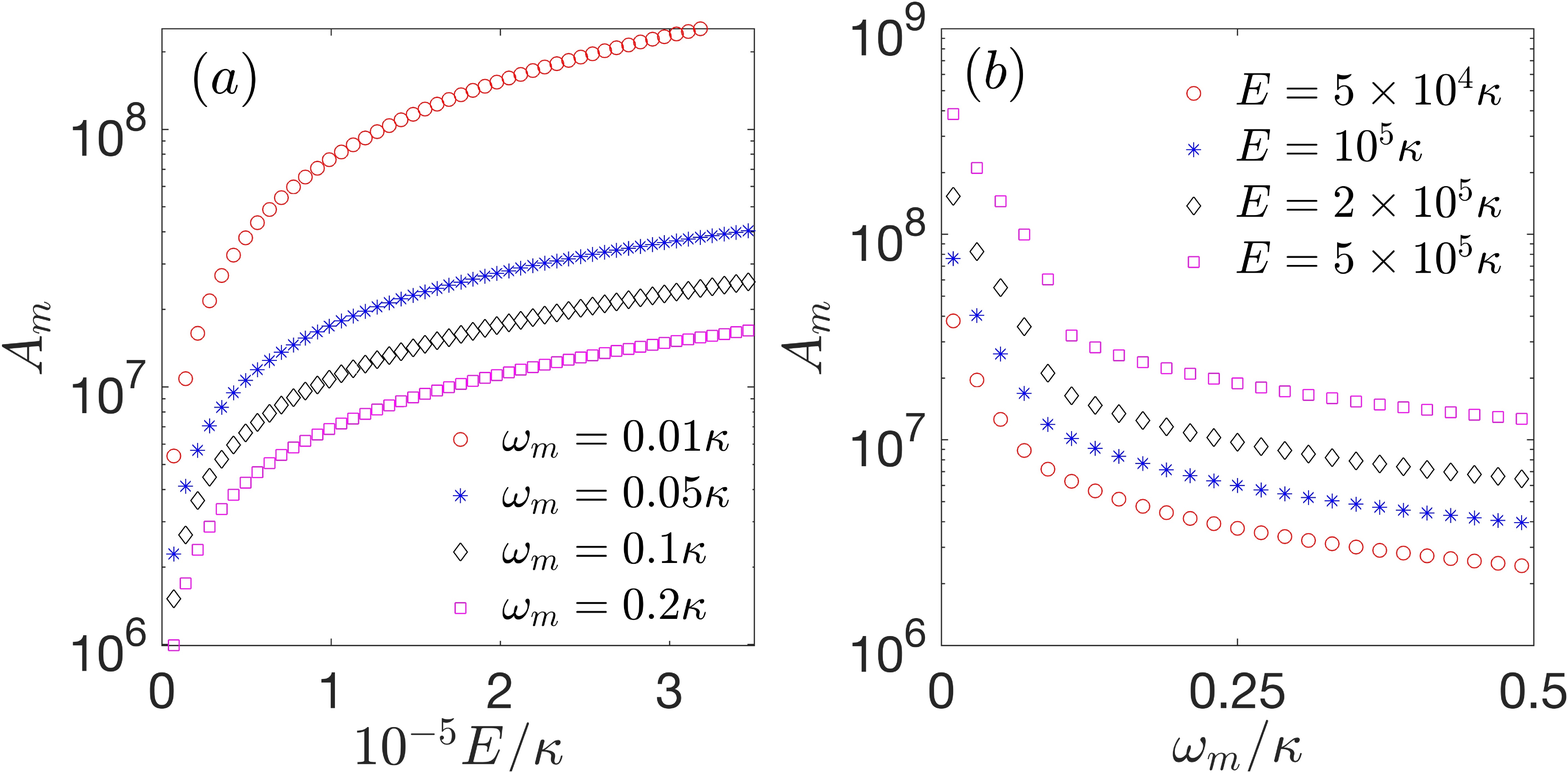}
	\caption{Realizable mechanical oscillation amplitudes $A_m$ due to varied drive amplitude (a) or different mechanical frequecnies (b). A large mechanical amplitude is possible with a high pump power or a small mechanical frequency. The fixed system parameters adopted in these simulation are $g_m=10^{-5}\kappa$, $\Delta_1=0$, $\Delta_2=-\omega_m$, and $Q=\omega_m/\gamma_m=1500$.}
	\label{fig6}
\end{figure}

So far, we have discussed on the figures-of-merit of the OMOFC generated by a specified two-tone drive on an OMS, which is operated by adjusting the pump-cavity coupling $\kappa_e$ to an unresolved sideband regime $\kappa>\omega_m$. In addition to locking the mechanical oscillation frequency to near the intrinsic one $\omega_m$, the comb span can be enlarged by times if one simply displaces the tones together to a proper detuning. A more obvious factor is the pump power $P=\hbar\omega_{1(2)}E^2/(2\kappa_e)$. In Fig. \ref{fig6}(a) we illustrate how the realized mechanical amplitude $A_m$ changes with the associated two-tone drive amplitude $E$. 
The realized amplitude $A_m$ for the different $\omega_m$ increases linearly with the drive amplitude [note that a logarithmic scale is used for the vertical axis of Fig. \ref{fig6}(a)]. If the system fabrication can enhance the optomechanical coupling $g_m$ to the level of $10^{-4}\kappa$, all systems with the used $\omega_m$ in Fig. \ref{fig6}(a) can realize the frequency comb line numbers $2g_mA_m/\omega_m>10000$ under the low pump powers, which achieve the dimensionless mechanical amplitude $A_m$ beyond $10^7$. Like the systems driven by a single-tone pump in unresolved sideband $\omega_m/\kappa<1$ \cite{ofc1}, those under a two-tone drive with the frequency difference $|\Delta_1-\Delta_2|$ close to $\omega_m$ encounter less problem of chaos when the pump power is extremely high, so the room for applying strong pump is big as long as the systems can withstand the heating by the intensified pulses in their cavities.

The mechanical frequency ratio $\omega_m/\kappa$ is another important factor to determine the comb span. Its relation with the realized mechanical amplitude is shown in Fig. \ref{fig6}(b). Especially, within the range of $\omega_m/\kappa<0.1\kappa$, the mechanical amplitude increases rapidly with this decreased ratio. For example, at the ratio $\omega_m/\kappa=0.05$, about $10500$ comb lines can be obtained with a relatively low drive amplitude $E=10^5\kappa$. More than $30500$ comb lines can be realized under the same pump power if the ratio $\omega_m/\kappa$ is decreased further to $0.01$.

\begin{figure*}[t]
	\centering
	\includegraphics[width=\linewidth]{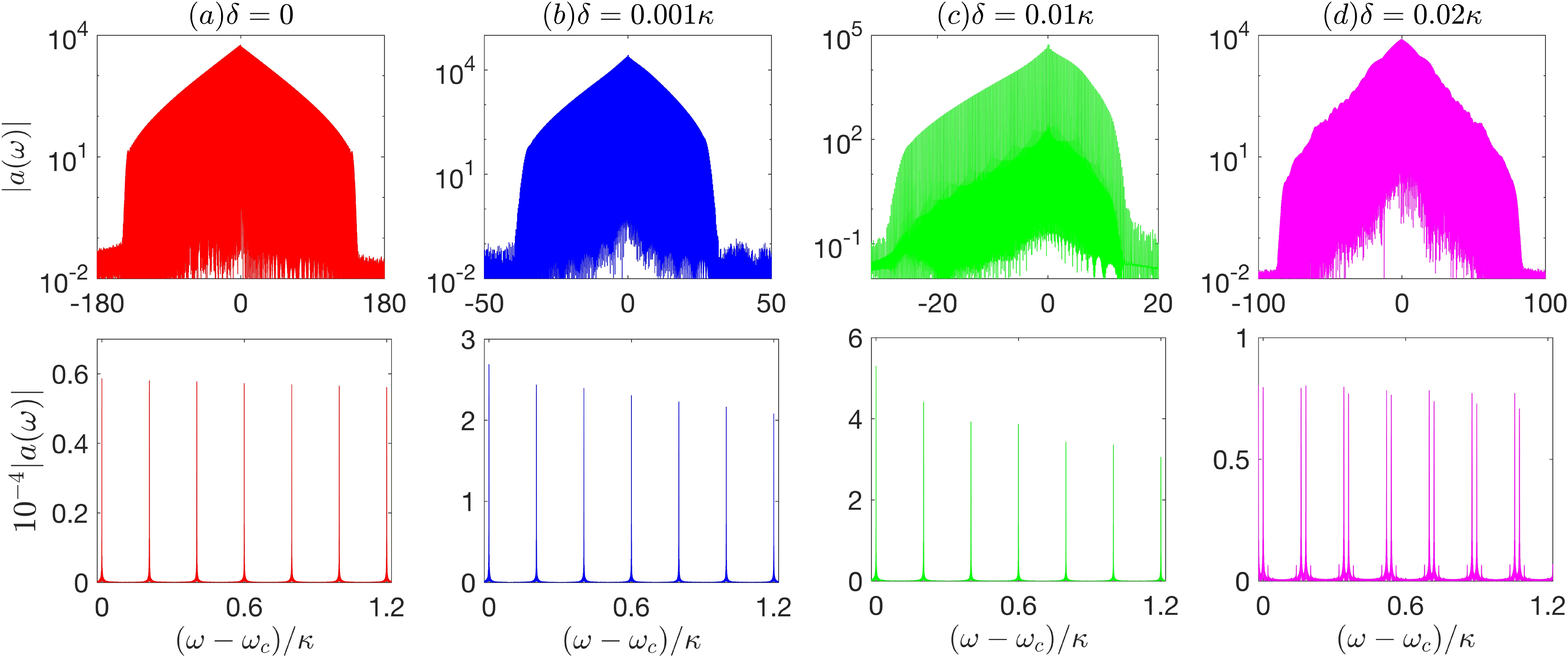}
	\caption{Possible consequences of imperfect frequency tone matching. The comb span or bandwidth is seen to be reduced by a larger error $\delta$ in Eq. (\ref{fm}). From (a) to (c), the distance between the comb teeth is $\omega_m+\delta$. After the error becomes as large as $\delta=0.02\kappa$, however, different sets of sideabands will be generated. These results are based on the system in Fig. \ref{fig3} by fixing the drive amplitude of the two tones at $\sqrt{2}E=4\times10^5\kappa$.}
	\label{fig7}
\end{figure*}

One method for a small ratio $\omega_m/\kappa$ is to have a higher rate $\kappa_e$ for the coupling between the pump and the cavity. However, it will also affect the other system parameters, especially the optomechanical coupling ratio $g_m/\kappa$. Then, it is better to have a mechanical mode with lower intrinsic frequency $\omega_m$, which can be realized by the fabrication to a different geometry of OMS. A smaller $\omega_m$ certainly implies a less spring restoring force proportional to $-\omega_m^2X_m$ and thus a larger oscillation amplitude $A_m$. From the view of photon-phonon interaction, a cavity photon with its energy $\hbar\omega_c$ will absorb (release) more phonons with a less energy $\hbar\omega_m$, so that it can be converted to more of those with the frequencies $\omega_c+n\omega_m$ ($\omega_c-n\omega_m$), where $n$ is an integer, to have a broader spectrum in Fig. \ref{fig1}(b). On the other hand, the cavity damping rate $\kappa$ can be regarded as the line width of the cavity field. In a good regime of generating OMOFCs, one such line width encompasses a large quantity of the sidebands distanced by the mechanical frequency $\omega_m$. 

For example, given the experimentally achievable parameters $\omega_m=0.05\kappa=5\pi$ MHz and $Q=1500$, together with an optomechanical coupling tuned to $g_m=2.3\times10^{-5}\kappa$ by the total cavity field damping rate $\kappa=2\pi\times50$ MHz, $9500$ comb lines can be obtained by the use of a two-tone pump with the detunings $\Delta_1=0$ and $\Delta_2=-0.05\kappa$. The required pump power corresponding to the drive amplitude $\sqrt{2}E=10^5\kappa$ is $32.1$ mW for the pump laser of $1550$ nm. Furthermore, the comb lines can be increased to $22500$ by simply displacing the detunings to $\Delta_1=-10\kappa, \Delta_2=-10.05\kappa$. When the pump power is decreased to $8.03$ mW, there will be $11150$ comb lines realized by the pump with the detunings $\Delta_1=-6.0\kappa$ and $\Delta_2=-6.05\kappa$. It should be noted that, for the convenience of reaching the final stability more quickly (see the discussion in Sec. 2.3), the mechanical quality factor $Q$ used for our simulations is low, and the better OMOFCs are expected to realize with a slight improvement on the quality of mechanical resonator. An even better improvement is with the optomechanical coupling strength $g_m$. If it can be increased by one order, a much enlarged comb span $2g_mA_m$ can be realized with the pump powers reduced by two orders.

\section{Imperfect drive-tone conditions}
\label{requirement}

In realistic situations, there exits an unavoidable error $\delta$ in Eq. (\ref{fm}) for the two tones of the driving field. This error $\delta$ could come from the fluctuation of the frequencies of a pump laser. To have a clear view of how this error would affect the generated OMOFC, we illustrate the effects from various errors in Fig. \ref{fig7}, which is also based on the exemplary process in Fig. \ref{fig3}. From Fig. \ref{fig7}(a) to Fig. \ref{fig7}(b), one sees that the generated OMOFC will only lose some sidebands to a smaller span $2g_mA_m$, if the error $\delta$ is within $10^{-3}\kappa$. This defect can be compensated by a parallel displacement of the two tones as in Fig. \ref{fig4}(b). For a setup with $\kappa\sim 2\pi\times 10$ MHz, the requirement on $\delta \sim 10^{-4}\kappa$ specifies that the distance between two drive tones should not deviate from the frequency $\omega_m$ by more than $2\pi\times 1$ kHz. It is feasible to apply acoustic-optic modulator (AOM) \cite{aom} or single-sideband modulator (SSB) \cite{ssb} to keep the error of two tones within the range. If a spectrum analysis of the mechanical oscillation is performed, one will see that the mechanical oscillation frequency under such a small error $|\delta|$ will be locked to the difference $|\Delta_1-\Delta_2|$ of the two drive tones. Up to the error $\delta=10^{-2}\kappa$ in Fig. \ref{fig7}(c), the mechanical oscillation frequency, as well as the distance between the individual sidebands, is still locked to the difference $|\Delta_1-\Delta_2|$, totally independent of the applied pump power. Therefore, the stability of the generated OMOFCs relies on how good we can control the frequency fluctuations of two drive tones.  

However, given a still larger error up to $\delta=2\times 10^{-2}\kappa$ for the exemplary system in Fig. \ref{fig7}, the dynamics of the system will become totally different. The effect of self-organized resonance will be lost, but the system never behaves as if it were under two independent drives. In this situation the two drive tones will realize other cooperative effects.
First of all, more mechanical frequency components, such as the fractional ones, will emerge while the peak of the mechanical spectrum $X_m(\omega)$ returns to the original mechanical frequency $\omega_m$ from $|\Delta_1-\Delta_2|$. The corresponding cavity field sidebands will therefore split into a number of different sets; the two main sets will be shifted to the frequency positions $\Delta_1+n\omega_m$ and $\Delta_2+n\omega_m$ ($n$ are integers), respectively, as shown in Fig. \ref{fig7}(d). By a further variation of the error $\delta$, the system will enter chaos or other complicated quasi-periodic oscillations. The dynamical patterns in the regimes of unmatched pump tones with $\delta\neq 0$ are determined by the ratio $\omega_m/\delta$; see more details in Ref. \cite{step}.

\section{Conclusion}
\label{conclusion}
The comb span of OMOFCs generated by an OMS is simply proportional to the mechanical oscillation amplitude it can possibly realize.
We have investigated an effective approach to add up the mechanical energy of oscillation for the OMS, so that it can realize broadband OMOFCs under feasible conditions. It is to operate the OMS in the regime of unresolved sideband and pump it by a two-tone field with the frequency difference of the drive tones being close to its intrinsic mechanical frequency. Then the system will be in a nonlinear resonance that quickly enlarges the mechanical oscillation amplitude and, purely through the intrinsic mechanical damping, it will stabilize to a final mechanical oscillation of significantly enhanced amplitude. Such enhanced mechanical oscillation amplitude can be increased further simply by choosing the proper detunings of the drive tones. A better operation point in the parameter space of the system will continually improve the generated OMOFC. With the experimentally feasible setups based on the mechanism, frequency comb lines of more than $10^4$ can be realized under pump powers in the order of mW. A very useful feature of the mentioned nonlinear resonance is that the mechanical oscillation frequency can be well locked to the frequency difference of the two drive tones. The resulting OMOFC pulse repetition rate and comb teeth spacing can be thus locked without any change by the pump power. 
It is therefore possible to achieve a good stability of the generated OMOFC by controlling the frequency fluctuations of two drive tones. The currently illustrated scenario, which is implementable by a specified two-tone pump, may reduce the distance to the 
real applications of OMOFCs.  

\vspace{-0cm}
\renewcommand{\theequation}{A-\arabic{equation}}
% redefine the command that creates the equation no.
\setcounter{equation}{0}  % reset counter 
\section*{Appendix}
Here we provide a derivation of the exact sideband amplitudes for the evolved cavity field $a(t)$ under a single-tone drive. After substituting the stabilized 
mechanical oscillation in Eq. (\ref{m}) in to the first equation of Eq. (\ref{dy}), which has been reduced to the situation of a single-tone drive, we obtain the following effective equation about the cavity field:
\begin{align}
	\dot{a}&=-\kappa a+ig_mA_m\cos(\Omega_mt)a+ig_md_ma+Ee^{i\Delta t}.
	\label{dy-c}
\end{align}
By a transformation $a(t)=e^{i\Delta t}e^{ig_mA_m\sin(\Omega_m t)/\Omega_m}a_1(t)$, the above equation will be reduced to 
\begin{align}
	\dot{a}_1&=-\kappa a_1-i\Delta a_1+ig_md_ma_1+Ee^{-ig_mA_m\sin(\Omega_m t)/\Omega_m}.
	\label{dy-c1}
\end{align}
Then, we make use of the generating function of the Bessel functions, $e^{(z/2)(x-1/x)}=\sum_{n=-\infty}^{\infty}J_n(z)x^n$, where 
$z=-gA_m/\Omega_m$ and $x=e^{i\Omega_mt}$, in the drive term of Eq. (\ref{dy-c1}) and solve it as a linear differential equation.
The solution is
\begin{align}
	a_1(t)=\sum_{n=-\infty}^{\infty}E\frac{J_n(-g_mA_m/\Omega_m)}{in\Omega_m+\kappa+i(g_md_m-\Delta)}e^{in\Omega_mt}.
	\label{sc}
\end{align} 
Using the generating function of the Bessel functions again, we obtain
\begin{align}
	a(t)&=\sum_{l=-\infty}^{\infty}\sum_{n=-\infty}^{\infty}E\frac{J_l(g_mA_m/\Omega_m)J_n(-g_mA_m/\Omega_m)}{in\Omega_m+\kappa+i(g_md_m-\Delta)}\nonumber\\
	& \times e^{i\Delta t+(l+n)\Omega_mt},
	\label{sc0}
\end{align} 
the coefficients of which can be combined to the form in Eq. (\ref{term}).

\vspace{0.4cm}

	\begin{acknowledgments}
	This work was supported by National Natural Science Foundation of China (Grant No. 12374348), and ANID Fondecyt
	Regular (1221250). The authors thank Dr. Ming Li for helpful discussions.
\end{acknowledgments}

\end{document}